\documentclass[11pt,hyper]{JHEP3}
\usepackage{amssymb,amsmath,amsfonts,cite,multirow,array}
\usepackage{graphicx}
\usepackage{dcolumn}
\usepackage{bm}
\usepackage{setspace}
\usepackage[normalem]{ulem}
\usepackage{relsize}
\usepackage{cancel}
\usepackage{verbatim}

\newcommand{\del}{\partial}

\title{Lifshitz Superfluid Hydrodynamics}

\author{Shira Chapman, Carlos Hoyos, Yaron Oz\\
Raymond and Beverly Sackler School of Physics and Astronomy, Tel-Aviv University, Tel-Aviv 69978, Israel\\
{\it E-mail:} \email{shirator@post.tau.ac.il, choyos@post.tau.ac.il, yaronoz@post.tau.ac.il}
}

\abstract{
We construct the first order hydrodynamics of quantum critical points with Lifshitz scaling and a spontaneously broken symmetry.
The fluid is described by a combination of two flows, a normal component
that carries entropy and a super-flow which has zero viscosity and carries no entropy.
We analyze the new transport effects allowed by the lack of boost invariance and constrain them by
the local second law of thermodynamics. Imposing time-reversal invariance, we find  eight new parity even transport coefficients.
The formulation is applicable, in general, to any superfluid/superconductor with an explicit breaking of boost symmetry, in
particular to high $T_c$ superconductors.
We discuss possible experimental signatures.
}


\begin{document}





\section{Introduction}

A phase transition at zero temperature may occur as the ground state of a many-body system is changed by tuning an external parameter. The boundary between the two phases is a quantum critical point \cite{Coleman:2005,Sachdev:2011}, characterized by a `Lifshitz' scaling symmetry
$t\to \lambda^z t, \ \ x^i\to \lambda x^i$,
where $t$ is time and $x^i$, $i=1,\dots, d$ are space coordinates. The number $z$ is the dynamical critical exponent. For an ordinary relativistic conformal field theory $z=1$, but for general systems its value can be arbitrary. Quantum critical points are believed to
underlie  the exotic properties of heavy fermion compounds and other materials including high $T_c$ superconductors \cite{2008NatPh,Si:2011hh}.

\begin{figure}[!ht]
	\begin{center}
	\includegraphics[width=4in]{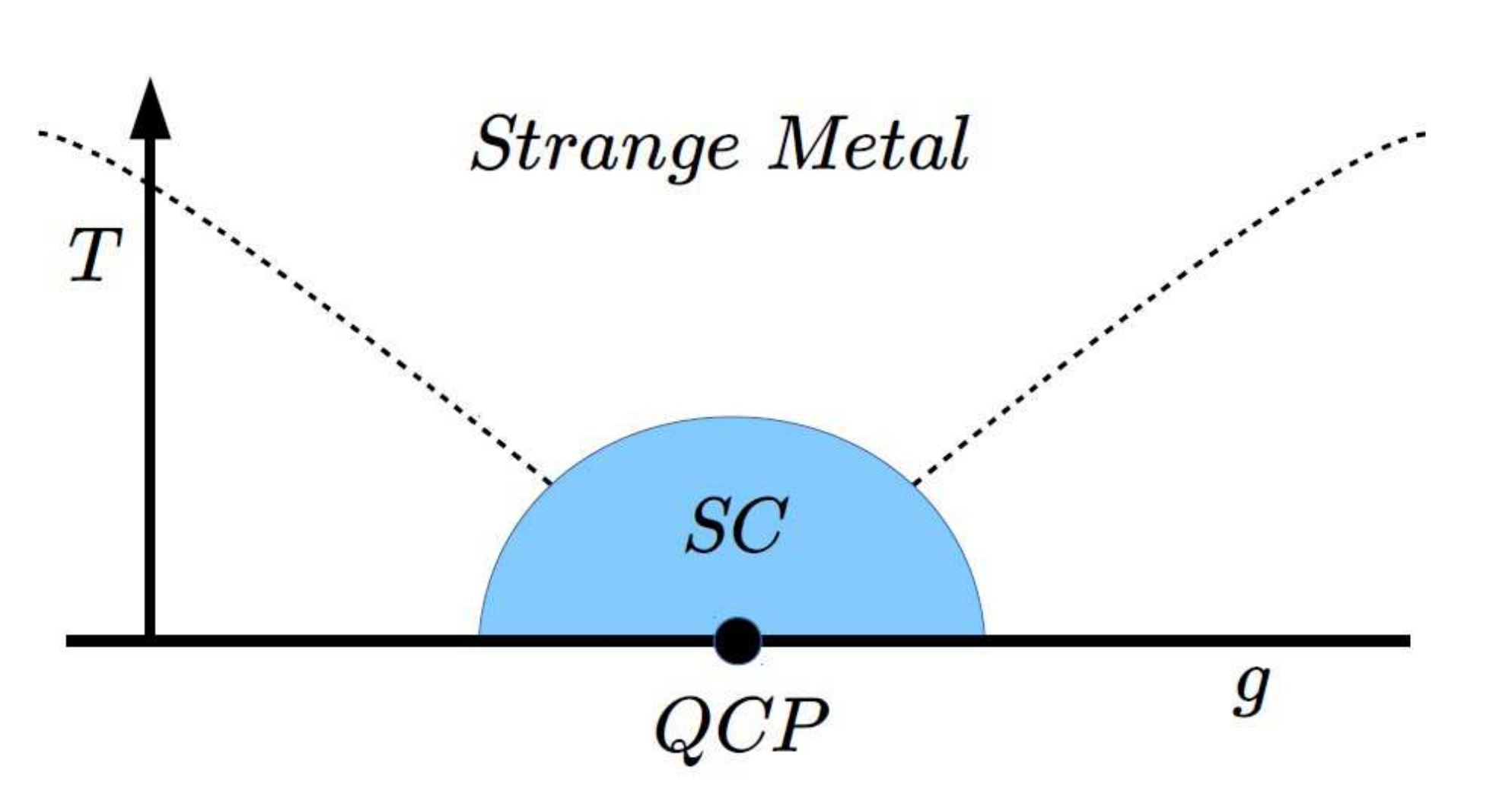}
	\caption{The phase diagram with a quantum critical point (QCP) hidden under the superconducting dome. $T$ is the temperature and $g$ is a quantum tuning parameter (e.g. doping, magnetic field). }\label{qcpphase}
	\end{center}
\end{figure}

At non-zero temperature, the low energy behavior of a quantum critical system may be well described by hydrodynamics.
A hydrodynamic description is valid when the characteristic length of thermal fluctuations $\ell_T \sim 1/T^{1/z}$ is much smaller than the correlation length $\ell_c\gg\ell_T$. This is the case in a V-shaped region of the phase diagram, and includes part of the superconducting dome where a symmetry
is spontaneously broken, as in figure ~\ref{qcpphase}. If the size of the system $L$ is smaller than the correlation length then deviations from criticality will be unimportant, but in the hydrodynamic approximation we also demand that gradients are much smaller than the temperature $\ell_c\gg L \gg\ell_T$.

The hydrodynamics of
a neutral fluid with Lifshitz symmetry has been constructed in \cite{Hoyos:2013eza}, and that of a charged Lifshitz fluid in \cite{Hoyos:2013qna}.
As noted above, Lifshitz scaling is potentially relevant in phases where a symmetry (global or local) is  spontaneously broken and
is interesting, for instance, in the study of high $T_c$ superconductors. In the superfluid/superconducting Lifshitz phase one should
consider the formation of a condensate. This leads to a hydrodynamic framework where the Lifshitz superfluid flow is a combination of two motions, a normal fluid component that carries entropy and a super-flow component (the gradient of the condensate phase $\partial_\mu \phi$) that has zero viscosity and carries no entropy. This is the case both for relativistic (see e.g. \cite{Bhattacharya:2011eea,Bhattacharya:2011tra,Neiman:2011mj}) and non-relativistic superfluids \cite{Landau,Putterman}. For a superconductor one considers the gauge-invariant combination of the gradient of the condensate phase and the external gauge field
$\xi_\mu=\partial_\mu \phi-A_\mu$.

The aim of this paper is to construct the hydrodynamics of a Lifshitz superfluid and superconductor.
The Lifshitz superfluid is characterized by the breaking of boost invariance (Lorentzian or Galilean). In the relativistic theory the breaking is manifest, since boosts mix space and time coordinates and the scaling symmetry introduces an anisotropy between them. We realize the breaking of Galilean invariance by taking the non-relativistic limit of a theory with broken Lorentz invariance. The currents associated to Lorentz symmetries are $J^{\mu\alpha\beta}=x^\alpha T^{\mu\beta}-x^\beta T^{\mu\alpha}$ (see e.g. Chapter 7 of \cite{weinberg}). The divergence of these Lorentz currents is given by:
\begin{equation}
\partial_\mu J^{\mu\alpha\beta} = T^{\alpha\beta}-T^{\beta\alpha}.
\end{equation}
Therefore, if the theory is Lorentz-invariant the energy-momentum tensor is symmetric.
We incorporate the breaking, without
introducing new hydrodynamic variables,\footnote{We do not add an external time direction vector to the system. This is very similar to the way in which the collective description of phonons does not require an external vector representing the lattice structure.} by allowing an antisymmetric contribution to the fluid stress-energy tensor. We define the local notion of the time direction as the time direction in the local rest frame of the normal fluid, thus the Lorentz/Galilean symmetry group is
broken  only in the direction of the normal-fluid velocity.\footnote{A possible generalization of this work would be to consider what happens when the symmetry is broken in a direction which is a linear combination of the normal and the superfluid velocities.}
We consider both relativistic and non-relativistic velocities.

We analyze new effects allowed by the lack of boost invariance in a parity preserving, time reversal invariant, superfluid and constrain them by the local second law of thermodynamics.
We find eight new transport coefficients satisfying Onsager's relations, associated with flows proportional to the acceleration of the normal fluid.
These transport coefficients are not present in a boost invariant superfluid and introduce new measurable effects. Our results are valid for any number of spatial dimensions $d\geq 2$.



The paper is organized as follows: in \S \ref{lifsf} we generalize the constitutive relations of a relativistic Lifshitz fluid to a superfluid. We study dissipative terms to first order in the derivative expansion and derive the constraints imposed by the local increase of the entropy density. In \S \ref{nrsf} we take the
non-relativistic limit and obtain the hydrodynamic equations of non-relativistic Lifshitz superfluid. In \S \ref{discus} we compare Lifshitz and non-Lifshitz
 superfluids and discuss possible new physical effects due to the breaking of boost invariance. In appendix \S \ref{nrlimitappendix} we present details of the non-relativistic limit of the Lifshitz superfluid.

\section{Lifshitz Superfluid Hydrodynamics }\label{lifsf}

Our construction of the first order Lifshitz superfluid hydrodynamics relies on previous works on Lifshitz hydrodynamics \cite{Hoyos:2013eza,Hoyos:2013qna} and on superfluid hydrodynamics \cite{Bhattacharya:2011eea,Bhattacharya:2011tra,Neiman:2011mj}. We will use similar notations to the ones presented in those papers. The energy-momentum tensor and current:
\begin{equation}\label{constit-general}
\begin{split}
& T^{\mu\nu}=(\varepsilon_n+p)u^\mu u^\nu +p \eta^{\mu\nu}+ f \xi^\mu \xi^\nu + \pi^{\mu\nu}\\
& J^\mu= q_n u^\mu-f \xi^\mu + J_{diss}^\mu \ ,
\end{split}
\end{equation}
where $u^\mu$ is the velocity ($\eta_{\mu\nu} u^\mu u^\nu=-1$), $\varepsilon_n$ the energy and $q_n$ the charge density of the normal component of the fluid. $p$ is the pressure, $\xi^\mu$ is proportional to the gradient of the phase of the condensate and $f$ determines the superfluid charge density. For a theory with Lifshitz scaling invariance, the scaling dimensions of the temperature and the chemical potential depend on the dynamical exponent $[T]=[\mu]=z$. The speed of light is also taken as a quantity with non-zero scaling dimension $[c]=z-1$. Then, the scaling dimensions of the terms appearing in the ideal energy-momentum tensor and current are as follows:
\begin{equation}\label{scaling}
[u^\mu]=0, \ \ [\varepsilon_n]=[p]=d+z, \ \ [q_n]=d,\ \  [f]=d-z, \ \ [\xi^\mu]=z.
\end{equation}
Note that derivatives have all the same scaling dimension $[\partial_\mu]=1$.\footnote{Note that $\partial_0\equiv\frac{1}{c}\partial_t$.} In general, transport coefficients with a scaling dimension $\Delta$ will have a dependence on thermodynamic quantities of the form:
\begin{equation}\label{scaledependencerel}
\sim T^{\frac{\Delta}{z}} F\left(\frac{\mu}{T}, \frac{\xi^2}{T^2} \right),
\end{equation}
where $F$ is an arbitrary function.

Our basic assumption is that no new hydrodynamic variables are needed for the description when boost symmetry is broken, and the effect of the broken symmetry is encoded in the antisymmetric contribution to the fluid stress-energy tensor. The thermodynamic relations are:
\begin{equation}\label{eq:thermo}
\varepsilon_n+p=Ts + \mu q_n,  \ \ dp=sdT+q_n d\mu+\frac{f}{2}\,d(\xi^2)\ ,
\end{equation}
and the Josephson relation:
$$ \ \ u^\mu\xi_\mu=\mu+\mu_{\rm diss}\ .$$
$\mu_{\rm diss}, J_{diss}^\mu, \pi^{\mu\nu}$ are first order corrections in a derivative expansion. At the derivative order, one has to choose a frame, due to the ambiguity in the hydrodynamic description.
The frame will be specified by five frame fixing conditions (associated to $T$, $\mu$ and $u^\mu$). In principle the superfluid velocity has a microscopic definition as the gradient of the Goldstone phase and it is not free for us to redefine. We will actually work in a modified phase frame where $\xi$ differs from the phase gradient by dissipative terms and can be treated as a normal thermodynamic variable. The frame is fixed by changing $\xi\to\xi_0$ with the conditions:\footnote{Thermodynamic relations still take the form \eqref{eq:thermo} with $\xi\to \xi_0$.}
\begin{equation}\label{eq:frame}
\xi_0^\mu=\zeta^\mu-\mu u^\mu, \ \ J_{diss}^\mu=0, \ \ \pi^{\mu\nu}u_\mu u_\nu=0,
\end{equation}
where $\zeta^\mu$ is the transverse component (with respect to $u^\mu$) of the phase gradient. The first condition corresponds to the modified phase prescription, while the other two take us to the Clark-Putterman frame. In particular this frame is better suited for taking the non-relativistic limit. A thorough discussion of the different frames can be found in the original reference \cite{Bhattacharya:2011eea}.

The main innovation with respect to previous works is that in a Lifshitz superfluid we allow an antisymmetric part for the stress-energy tensor of the form:
\begin{equation}
\pi_A^{[\mu\nu]} = u^{[\mu}V_A^{\nu]}\ ,
\end{equation}
where $V^A_{\mu}$ is any vector that can be built from the fluid data and we have allowed for breaking of boost symmetry only, rotations must be kept intact.\footnote{More explicitly, the two conditions we impose are $T^{\mu\nu}u_\mu P_\nu^{\ \lambda} \neq T^{\mu\nu} P_\mu^{\ \lambda} u_\nu$ and $T^{\mu\nu}P_\mu^{\ \sigma} P_\nu^{\ \lambda} = T^{\mu\nu} P_\mu^{\ \lambda} P_\nu^{\ \sigma}$ \cite{Hoyos:2013eza,Hoyos:2013qna}.}
In the following we will use the local second law of thermodynamics to constrain the first derivative corrections to the constitutive relations and the Josephson equation.

\subsection{Ideal Hydrodynamics}

One may wonder why we did not consider at the ideal level an additional term in the stress-energy tensor of the form:
\begin{equation}\label{T_zero_corr}
T_A^{[\mu\nu]} = C u^{[\mu}\xi_0^{\nu]}\ .
\end{equation}
The reason is that such a term is not consistent with vanishing entropy production that is required at the ideal order.
The equation for the entropy production rate is constructed at zeroth order as follows:
$$
0=\partial_\mu T^{\mu\nu} u_\nu+\mu \partial_\mu J^\mu=-T\partial_\mu(s u^\mu)\ .
$$
We ask  whether it is possible to modify the entropy current, charge current, stress-energy tensor  and the thermal relations, such that the entropy production
rate vanishes.

By carefully considering all these possibilities one can show that a term of the form  \eqref{T_zero_corr} cannot be absorbed by the above redefinitions for any $C \neq 0$. The reason for this is the appearance of a $~\xi^\mu a_\mu$ term in the entropy production rate where the acceleration is $a_\mu \equiv u^\nu \del_\nu u_\mu$. This term cannot be compensated for by any of the redefinitions described above. We therefore conclude that \eqref{constit-general} are the most general constitutive relations allowed at zeroth order.

\subsection{The First Dissipative Order}

Consider next the first derivative order. Since we assumed that no new hydrodynamic variables are needed for our description of a superfluid with broken boost invariance, the full classification of scalars/vectors/tensors can be built from the one derivative fluid data that can be found in table (3) of \cite{Bhattacharya:2011tra}. The classification of tensors is according to the ``explicit'' breaking of Lorentz symmetry by the presence of a normal and superfluid velocities to an $SO(d-1)$ symmetry group (see \cite{Bhattacharya:2011tra} for details).

From the frame conditions \eqref{eq:frame} it follows that we can decompose the symmetric part of the stress-energy tensor as
\begin{equation}\label{pimn}
\pi^{(\mu\nu)} = (Q^\mu u^\nu + Q^\nu u^\mu) + \Pi P^{\mu\nu} + \Pi^{\mu\nu}_t\ ,
\end{equation}
where
\begin{equation}
Q^\mu u_\mu=0, \ \ \Pi^{\mu\nu}_t u_\nu=0, \ \ \Pi_t^{\mu\nu}P_{\mu\nu}=0\ .
\end{equation}


To describe the Lifshitz superfluid we add an antisymmetric part to the stress-energy  tensor of the form:
\begin{equation}\label{pia}
\pi_A^{[\mu\nu]} = u^{[\mu}V_A^{\nu]} \ ,
\end{equation}
where $V_A^{\mu}$ is any vector constructed from fluid data at first order in derivatives.
The stress-energy tensor then takes the form:
\begin{equation}
T^{\mu\nu} = (\varepsilon_n+p) u^\mu u^\nu +p \eta^{\mu\nu} + f \xi_0^\mu \xi_0^\nu+ \pi^{(\mu\nu)} + \pi_A^{[\mu\nu]}\ ,
\end{equation}
where all thermal functions are functions of $\mu, T, \xi_0$.

In section 3.2 of \cite{Bhattacharya:2011tra} a listing of all possible corrections to the entropy current was presented and their contributions to the divergence of the entropy current were evaluated. The analysis is left unchanged in our case.\footnote{\label{f5}
In an earlier version of this paper we relied on a slightly modified version of the arguments of \cite{Bhattacharya:2011tra} to claim that the entropy current should take its canonical form in parity even theories independent of the assumption of time reversal invariance. This result was corrected in a newer version of \cite{Bhattacharya:2011tra}. The form of the entropy current therefore only holds for time reversal invariant theories. We will only consider such theories in what follows.}

In the modified Clark-Putterman frame the entropy current takes the following form:
\begin{equation}
J^\mu_s
=s u^\mu - \frac{u_\nu}{T} \pi^{\mu\nu}
+\frac{f}{T}\mu_{diss}\zeta^\mu \ ,
\end{equation}
and its divergence is given by:\footnote{For a derivation see section 2.6 of \cite{Bhattacharya:2011tra}.}
\begin{equation}
\begin{split}
\del_\mu J^\mu_s = &
- \del_\mu \left(\frac{u_\nu}{T}\right) \pi^{\mu\nu}  + \mu_{diss} P^{\mu\nu} \del_\mu \left(\frac{f\zeta_\nu}{T}\right)\ .
\end{split}
\end{equation}
In the derivation we have used the curl free condition $\partial_\mu \xi_\nu-\partial_\nu\xi_\mu=0$, or in terms of $\xi_0$,
\begin{equation}\label{eq:rotat}
\partial_\mu\xi_{0\,\nu}-\partial_\nu\xi_{0\,\mu}=\partial_\mu(\mu_{diss} u_\nu)-\partial_\nu(\mu_{diss} u_\mu)\ .
\end{equation}

In the relativistic Clark-Putterman frame \eqref{eq:frame} the divergence of the entropy current is thus modified by a new term depending on $V_A^\mu$:
\begin{align}\label{div_entropy_our}
\begin{split}
\del_\mu J^\mu_s = &
- \frac{\left[ \Pi( \partial_\mu u^\mu)+ \Pi_t^{\mu\nu}\sigma_{\mu\nu}\right]}{T}
- \frac{Q_\mu}{T}\left[ a^\mu+ P^{\mu\nu}\frac{\partial_\nu T}{T} \right]
+\mu_{diss} P^{\mu\nu} \del_\mu \left(\frac{f\zeta_\nu}{T}\right)\\
& - \frac{V_{A\,\mu}}{2T}\left[ a^\mu -
P^{\mu\nu} \frac{\del_\nu T}{T} \right]\ ,
\end{split}
\end{align}
where $\sigma_{\alpha\beta} \equiv \del_{\langle \alpha} u_{\beta \rangle}\equiv P_\mu^\alpha P_\nu^\beta
(\partial_{(\alpha} u_{\beta)}-P_{\alpha\beta} (\del \cdot u)/d)$ is the shear tensor of the normal fluid velocity.
The right hand side of \eqref{div_entropy_our} can be written as a sum of three classes of
quadratic forms in one derivative data - scalars, vectors and tensors (under the group $SO(d-1)$ which leaves $u^\mu, \xi^\mu$ unchanged).
In principle all possible independent fluid data could enter in the dissipative corrections, but actually only those that appear explicitly in \eqref{div_entropy_our} contracted with the dissipative corrections to the currents are allowed. The reason is that the condition of positive entropy production rate translates into the requirement that equation \eqref{div_entropy_our} should be a positive sum of squares. If we denote by $x$ one of the factors that appear explicitly in \eqref{div_entropy_our} and we allow dissipative terms to contain both $x$ and other fluid data $y$ that does not appear explicitly, then in the divergence of the entropy current there will be quadratic terms of the form $x^2+cxy$ where $x$ and $y$ are independent pieces of fluid data. This cannot be positive definite unless $c=0$.\footnote{A general analysis like the one presented in \cite{Bhattacharya:2011tra} would require considering additional transport coefficients on the right hand side of \eqref{div_entropy_our} that can be recast in the form of total derivatives, and therefore reabsorbed by modifying the definition of the entropy current. We have checked that no such terms are allowed in our case for time reversal invariant theories, see also footnote~\ref{f5}.}


Let us now list the independent fluid data. For convenience we will define the vectors
\begin{equation}
b_\pm^\mu = a^\mu \pm P^{\mu\nu}\frac{\partial_\nu T}{T}\ .
\end{equation}
We have in the scalar sector 6 independent scalars. We choose 5 of them to be  
\begin{equation}\label{scalar_data}
\Sigma_0 = n_\nu b_-^\nu ,\ \
\Sigma_1 = n_\nu b_+^\nu ,\ \
\Sigma_2 = n^\mu \sigma_{\mu\nu} n^\nu,\ \
\Sigma_3 = \partial_\mu u^\mu,\ \
\Sigma_4 =T 
P^{\mu \nu} \partial_\mu \left(\frac{f~\zeta_\nu}{T} \right),
\end{equation}
where we have defined $n^\mu = \zeta^\mu/\zeta$ and we will also use the projector $\tilde P^{\mu\nu}=P^{\mu\nu}-n^\mu n^\nu$.
In the vector sector we have 5 independent vectors. We choose 3 of them to be 
\begin{equation}\label{vector_data}
{\cal W}_0^\mu =\tilde P^{\mu\nu} b_{-\,\nu},\ \
{\cal W}_1^\mu =\tilde P^{\mu\nu} b_{+\,\nu},\ \
{\cal W}_2^\mu =n^\beta \sigma_{\beta\alpha} \Tilde P^{\alpha\mu}.
\end{equation}
We also have two tensors available for our construction. We choose one of them to be
\begin{equation}
\begin{split}\label{tensor_data}
\mathcal{T}_{\mu\nu} = & \tilde P_\mu^\alpha \tilde P_\nu^\beta
\sigma_{\alpha\beta} \ .
\end{split}
\end{equation}
Note that any term involving the acceleration is in fact related to heat transfer by the transverse projection of the ideal order equations of motion $\del_\mu T^{\mu\nu} = 0$:
\begin{equation}\label{eq:acc}
\begin{split}
a^\mu \equiv u^\alpha\del_\alpha u^\mu =-P^{\mu\sigma}\left(\frac{\partial_\sigma T}{T}+
\frac{q_n T}{\varepsilon_n + p}\del_\sigma \frac{\mu}{T} \right)  - \zeta^\mu \frac{\del_\sigma\left(f\xi_0^\sigma\right)}{\varepsilon_n + p}\ .
\end{split}
\end{equation}
For a theory with Lifshitz scaling we can read the scaling dimensions of all these terms using \eqref{scaling}. They are:
\begin{equation}\label{dissscalingrel}
[\Sigma_{i\leq 3}]=[{\cal W}_i^\mu]=[\mathcal{T}_{\mu\nu}]=1, \ \ [\Sigma_4]=d+1.
\end{equation}


We can now write the most general allowed form for the first order corrections to the constitutive relations:\footnote{
Note that the orthogonal scalar basis for the tensor decomposition was chosen to be $P^{\mu\nu}, n^\mu n^\nu-P^{\mu\nu}/d$ rather then the more intuitive $\tilde P^{\mu\nu}, n^\mu n^\nu$. This is merely for numerical convenience in taking the non-relativistic limit. $\tilde {\cal T}^{\mu\nu} = {\cal T}^{\mu\nu} + \Sigma_2/(d-1) \cdot \tilde  P^{\mu\nu}$ is the traceless part of ${\cal T}^{\mu\nu}$.}
\begin{equation}\label{general_constit1}
\begin{split}
\mu_{diss} = \ & 
 \sum_{i=0}^4 \mu_i \Sigma_i+\cdots \\
- \Pi = \ & 
 \sum_{i=0}^4 \pi_i \Sigma_i+\cdots \\
 - Q^\mu = & 
 n^\mu\sum_{i=0}^4 Q^{(s)}_i \Sigma_i
+ 
 \sum_{i=0}^2 Q^{(v)}_i {\cal W}_i^\mu+\cdots \\
 - V_A^\mu = & \ 2 
 n^\mu\sum_{i=0}^4 V^{(s)}_i \Sigma_i
+ 2 
 \sum_{i=0}^2 V^{(v)}_i {\cal W}_i^\mu+\cdots \\
- \Pi_t^{\mu\nu} = \ & 
\left(n^{\mu} n^{\nu} - \frac{1}{d}P^{\mu\nu}\right)\sum_{i=0}^4 P_i \Sigma_i
  +
   \sum_{i=0}^2 E_i {\cal W}_i^{(\mu} n^{\nu)} \\
&+  \tau 
 \left[2 \tilde {\cal T}^{\mu\nu}  + \frac{2\Sigma_2}{d-1} \left(n^{\mu} n^{\nu} - \frac{1}{d}P^{\mu\nu}\right) \right] + \cdots\ , 
\end{split}
\end{equation}
where the factors of $T$ were inserted for later convenience. The ``$\cdots$'' stand for the additional scalars/vectors/tensors completing our basis \eqref{scalar_data}, \eqref{vector_data}, \eqref{tensor_data} of independent fluid data to 6 scalars, 5 vectors and 2 independent tensors. Their explicit form will not be needed in what follows since, as explained below \eqref{div_entropy_our}, they are ruled out by positivity of the entropy production rate. We can determine the scaling dimensions of the coefficients from \eqref{dissscalingrel} and using the fact that $[T^{\mu\nu}]=d+z$ and $[\mu]=z$. We get:
\begin{equation}
\begin{split}
& [\mu_i]=[\pi_4]=[ Q_4^{(s)}]=[V_4^{(s)}]=[P_4]=z-1, \ \  i\leq 3,\ \  
[\mu_4]=z-(d+1) 
\ ,\\
&[\pi_i]=[ Q_i^{(s)}]=[V_i^{(s)}]=[ Q_i^{(v)}]=[V_i^{(v)}]=[P_i]=[E_i]=[\tau]=d+z-1
, \ \ i\leq 3\ .
\end{split}
\end{equation}
This partially fixes the dependence of the transport coefficients on the thermodynamic variables to be general functions of the form \eqref{scaledependencerel}.

Plugging \eqref{general_constit1} into \eqref{div_entropy_our} we can write the divergence of the entropy current as a linear sum of three different quadratic forms (involving the tensor terms, vector terms and scalar terms respectively):
\begin{equation} \label{divcur}
\partial_\mu J^\mu_s= \frac{1}{T}(Q_s+ Q_V+Q_T) \ , 
\end{equation}
where,
\begin{equation}
\begin{split}
Q_T=& \ 2\tau\mathcal{T}^2\ ,\\
Q_V=& \ \sum_{i=0\dots 2}\left[V_i^{(v)} {\cal W}_i \cdot {\cal W}_0 + Q_i^{(v)} {\cal W}_i \cdot {\cal W}_1 + E_i {\cal W}_i  \cdot {\cal W}_2 \right]\ ,\\
Q_S=& \ \sum_{i=0\dots 4} \left[ V_i^{(s)}  \Sigma_i \Sigma_0 + Q_i^{(s)} \Sigma_i \Sigma_1
  +  P_i \Sigma_i \Sigma_2 +  \pi_i \Sigma_i \Sigma_3 +  \mu_i \Sigma_i \Sigma_4 \right]\ .\\
\end{split}
\end{equation}
All these coefficients can be rearranged into matrices as follows:
$$Q_V= \begin{bmatrix}{\cal W}_0 & {\cal W}_1 & {\cal W}_2 \end{bmatrix}
\begin{bmatrix}  V^{(v)}_0 & Q^{(v)}_0 &  E_0 \\ V^{(v)}_1 & Q^{(v)}_1 &  E_1 \\ V^{(v)}_2 & Q^{(v)}_2 & E_2  \end{bmatrix}
\begin{bmatrix}{\cal W}_0 \\ {\cal W}_1 \\ {\cal W}_2 \end{bmatrix} \equiv \vec {\cal W}^T \left[\vec V^{(v)}, \vec Q^{(v)} , \vec E \right] \vec {\cal W} \ ,$$
$$ Q_S=  \begin{bmatrix}\Sigma_0 & \Sigma_1 & \Sigma_2 & \Sigma_3 & \Sigma_4 \end{bmatrix}
\begin{bmatrix} V^{(s)}_0 & Q^{(s)}_0 & P_0 & \pi_0 & \mu_0 \\
                V^{(s)}_1 & Q^{(s)}_1 & P_1 & \pi_1 & \mu_1 \\
                V^{(s)}_2 & Q^{(s)}_2 & P_2 & \pi_2 & \mu_2 \\
                V^{(s)}_3 & Q^{(s)}_3 & P_3 & \pi_3 & \mu_3 \\
                V^{(s)}_4 & Q^{(s)}_4 & P_4 & \pi_4 & \mu_4 \\
                      \end{bmatrix}
\begin{bmatrix} \Sigma_0 \\ \Sigma_1 \\ \Sigma_2 \\ \Sigma_3 \\ \Sigma_4 \end{bmatrix} \equiv
\vec \Sigma^T \left[\vec V^{(s)}, \vec Q^{(s)}, \vec P, \vec \pi, \vec \mu\right] \vec \Sigma \ .$$

To have a positive entropy production rate these matrices are required to be positive semi-definite. We can split the matrices into symmetric and antisymmetric parts. The antisymmetric part will vanish if we impose Onsager's relations that follows from the requirement of time reversal invariance. 
The matrices can then be diagonalized through an orthogonal transformation, and the positivity condition becomes a condition on the eigenvalues. In the usual relativistic superfluid, the first row and column of the matrices above would be absent.  The number of transport coefficients is then 14 for the non-Lifshitz case compared to 22 in the Lifshitz case. We see that we have 8 new transport coefficient introduced by the lack of boost invariance.
Since we do not allow parity breaking, all the transport coefficients are parity even.


\section{Non-Relativistic Lifshitz Superfluid}\label{nrsf}

In the non-relativistic limit $c\to \infty$ the relativistic superfluid equations should become Landau's two-fluid hydrodynamic description \cite{Landau,Putterman}. Although this is a straightforward exercise, the details are a bit cumbersome and we relegate them to appendix \ref{nrlimitappendix}. We will just present here the result of taking the limit and analyze it.

In Landau's description, there is a mass density $\rho$ which is the sum of the contributions from the normal ($\rho_n$) and the superfluid ($\rho_s$) components. The motions of the normal and superfluid components are independent and  are described by different velocities $v^i$ and $v_s^i$, respectively.
 An important  role is played by the 'counterflow', the difference between the normal and superfluid velocities, that we define as $w^i=v_s^i-v^i$. Note that the counterflow is invariant under Galilean boosts.

In terms of these variables, the  hydrodynamic equations become:
\begin{equation}
\begin{split}
&\partial_t (\rho_n+\rho_s)+\partial_i(\rho_n v^i+\rho_s v_s^i)=0 \ ,\\
&\partial_t(\rho_n v^i+\rho_s v_s^i)+\partial_k(\rho_n v^i v^k+\rho_s v_s^i v_s^k)+\partial^ip+\nu^i=0 \ ,\\
&T(\partial_t s+\partial_i j_s^i)+\nu_s=0\ ,\\
&\partial_t v_s^i+\partial^i\left(\frac{v_s^2}{2}+\mu+H \right)=0\ ,
\end{split}
\end{equation}
where $j_s^i$ is the entropy current and $\nu_s$ and $\nu_i$ are dissipative terms that we will specify below.  The chemical potential $\mu$ is  not the same variable as in the relativistic theory, but they are closely related, the details can be found in the appendix. $H = c \mu_{\rm diss}$ is a dissipative term. The equations can be identified as the continuity equation for the mass, the generalization of the Navier-Stokes equations for the superfluid and the equation for the entropy current, respectively. The last equation implies that $v_s^i$ is the gradient of a scalar and is often referred to as the law of potential flow.

Assuming Lifshitz symmetry, the scaling dimensions of the thermodynamic variables and velocities should be the following:\footnote{The scaling dimension of the chemical potential is different from the one in the relativistic theory because it is associated to the {\em mass} density, so it is divided by a factor with scaling dimension $[m]=2-z$ relative to the relativistic case.}
\begin{align}\label{scalingNR}
\begin{split}
&[\partial_t]=[T]=z, \ [\partial_i]=1, \ [v^i]=[v_s^i]=z-1,\ [\mu]=2(z-1)\ , \\
 &[p]=z+d, \ [\rho_n]=[\rho_s]=d+2-z, \ [s]=d\ .
 \end{split}
\end{align}
The scaling dimensions of the dissipative terms are then fixed to be:
\begin{equation}\label{scalingDis}
[H]=2(z-1), \ \ [\nu^i]=z+d+1, \ \ [\nu_s]=2z+d\ .
\end{equation}
A coefficient with scaling dimension $\Delta$ will have the following general dependence on the thermodynamic variables:
\begin{equation}\label{scaledependenceNR}
\sim T^{\frac{\Delta}{z}}F\left(\frac{\mu}{T^{\frac{2(z-1)}{z}}},\frac{w^2}{T^{\frac{2(z-1)}{z}}} \right)\ .
\end{equation}

In the non-relativistic limit the qualitatively new contributions to the superfluid motion come from the terms that contain $\Sigma_0+\Sigma_1$ and ${\cal W}_0^\mu+{\cal W}_1^\mu$ (that is, the acceleration $a^\mu$).
In order to simplify the notation we define
the projector:
\begin{equation}
P_w^{ij}=\delta^{ij}-\frac{w^iw^j}{w^2}\ .
\end{equation}
Then, the new contributions take the form:
\begin{equation}
\Sigma_0+\Sigma_1 \sim \frac{w^i}{w}D_tv_i , \ \ {\cal W}_0^i+{\cal W}_1^i\sim P_w^{ij}D_t v_i+\cdots\ .
\end{equation}
Where we have defined the `material derivative' $D_t$ as:
\begin{equation}
D_t v^i=(\partial_t+v^k\partial_k)v_i\ .
\end{equation}
We see that the qualitatively new dissipative terms contain the acceleration (projected with respect to the counterflow direction), which are absent in non-Lifshitz superfluids.

The dissipative terms contain 14 transport coefficients that coincide with the non-Lifshitz superfluid and that can be found in the literature (see for instance \cite{Bhattacharya:2011eea,Putterman}). In the Lifshitz fluid there are 8 additional transport coefficients that satisfy Onsager's relations. In order to find the non-relativistic form of the dissipative terms we expand the relativistic terms in powers of the speed of light:
\begin{equation}
\pi^{\mu\nu}=\sum_n \frac{1}{c^n} \pi_{(n)}^{\mu\nu}\ ,
\end{equation}
where $n\geq -1$ depending on the component we pick. The entropy current and non-relativistic dissipative terms are:
\begin{equation}
\begin{split}
j_s^i &= s v^i+\frac{1}{T}\left(\pi_{(1)}^{i0}-\pi_{(0)}^{ik} v_k\right)\ ,\\
\nu_s &= \left(\pi_{(1)}^{i0}-\pi_{(0)}^{ik} v_k \right)\frac{\partial_i T}{T} +\pi_{(-1)}^{0i}\partial_t v_i+\pi_{(0)}^{ki}\partial_k v_i
+H \del_i(\rho_s w^i)\ ,\\
\nu^i&=\partial_t\pi^{0i}_{(-1)}+\partial_k \pi^{ki}_{(0)}\ .
\end{split}
\end{equation}
Note that the relativistic description is covariant, and the same must be true in the non-relativistic limit, but the term proportional to $\pi_{(-1)}^{0i}$ is not covariant. In order to recover covariance, $\pi^{ij}$ must contain an asymmetric term:
\begin{equation}
 \pi^{ij}_{(0)}=v^i \pi_{(-1)}^{0j}+\pi_{T\,(0)}^{(ij)}\ ,
\end{equation}
where $\pi_{T\,(0)}^{(ij)}$ is a symmetric contribution originating from the transverse terms in $\pi^{\mu\nu}$. The origin of the asymmetric term can be understood from \eqref{pimn} and \eqref{pia}. The leading term appears in $\pi_{(-1)}^{0i}$ through the combination
$\pi^{0i} \sim V_A^i+Q^i$, while the combination appearing in $\pi^{i0}\sim -V_A^i +Q^i$ is suppressed. Both combinations appear in $\pi^{ij}$, but in the non-relativistic limit only one survives at the right order $\pi^{ij}\sim \frac{v^i}{c}( V_A^j+Q^j)$. Therefore, we can rewrite the dissipative terms as:
\begin{align}\label{nus}
\begin{split}
\nu_s &= \left(\pi_{(1)}^{i0}-\pi_{(0)}^{ik} v_k \right)\frac{\partial_i T}{T} +\pi_{(-1)}^{0i}D_t v_i+\pi_{T\,(0)}^{(ki)}\partial_k v_i
  +H \del_i(\rho_s w^i)\ ,\\
\nu^i&=\partial_t\pi^{0i}_{(-1)}+\partial_k \left(v^k\pi^{0i}_{(-1)}\right) +\partial_k \pi^{(ki)}_{T\,(0)}\ .
\end{split}
\end{align}
The 5 independent scalars of the relativistic superfluid map to:\footnote{The map involves different linear combinations of the $\Sigma_i$ scalars in the non-relativistic limit.}
\begin{equation}\label{scalar_dataNR}
S_0 = w^i D_t v_i ,\ \
S_1 = w^i\partial_i T ,\ \
S_2 = w^i w^j \sigma_{ij},\ \
S_3 = \partial_i v^i,\ \
S_4 = \partial_i \left(\rho_sw^i\right),
\end{equation}
where $\sigma_{ij}=\partial_{(i} v_{j)}-\frac{1}{d}\delta_{ij}\partial_k v^k$ is the shear tensor. The 3 independent vectors map to:
\begin{equation}\label{vector_dataNR}
W_0^i = P_w^{ij} D_t v_j,\ \
W_1^i =  P_w^{ij}\partial_j T,\ \
W_2^i =w^jP_w^{ik} \sigma_{jk}.
\end{equation}
The independent tensor maps to:
\begin{equation}
\begin{split}\label{tensor_dataNR}
\widetilde{T}^{ij} = & P_w^{ik} P_w^{jl}
\sigma_{kl} \ .
\end{split}
\end{equation}
Compared to the non-Lifshitz fluid, $S_0$ and $W_0^i$ are new allowed terms. Using \eqref{scalingNR} we can determine the scaling dimensions of these terms in a theory with Lifshitz symmetry:
\begin{align}
\begin{split}
[S_0]=[S_2]=3z-2,\ [S_1]=2z,\ [S_3]=z, \ [S_4]=d+2\ ,\\ \label{dissscalingNR}
[W_0^i]=[W_2^i]=2z-1,\ [W_1^i]=z+1,\ [\widetilde{T}^{ij}]=z\ .
\end{split}
\end{align}
In this basis we can expand the dissipative terms as:
\begin{align}\label{constit_nr2}
\begin{split}
&H = -\sum_{A=0}^4 A^s_A S_A\ ,\\
&\pi_{(1)}^{i0}-\pi_{(0)}^{ik} v_k = -w^i\sum_{A=0}^4 B^s_A S_A-\sum_{A=0}^2 B^v_A W_A^i\ ,\\
&\pi_{(-1)}^{0i} = -w^i\sum_{A=0}^4 C^s_A S_A-\sum_{A=0}^2 C^v_A W_A^i\ ,\\
&\pi_{T\,(0)}^{(ij)} =- w^iw^j\sum_{A=0}^4 D^s_A S_A-P_w^{ij}\sum_{A=0}^4 E^s_A S_A-\sum_{A=0}^2 D^v_A w^{(i} W_A^{j)}-\eta\tilde{T}^{ij}\ .
\end{split}
\end{align}
Possible powers of the speed of light have been absorbed in the definition of the transport coefficients, that now depend on the thermodynamic variables $T$, $\mu$ and $w^2$.
The dissipative terms appearing in the equation for the entropy should be negative semi-definite:
\begin{equation}
\nu_s\leq 0\ .
\end{equation}
In order for this to happen, we should be able to group them in complete squares, leading to the same kind of positivity conditions that we derived in the relativistic case.

With the help of \eqref{scalingNR}, \eqref{scalingDis},
\eqref{nus}, \eqref{dissscalingNR} and \eqref{constit_nr2}, we can derive the following scaling dimensions for the transport coefficients:
\begin{align}
\begin{split}
&[A_0]=[A_2]=-z, \ [A_1]=-2, \ [A_3]=z-2,\ [A_4]=2(z-2)-d\ , \\
&[B_0^s]=[B_2^s]=d-2(z-1),\ [B_1^s]=d-z,\ [B_3^s]=d,\ [B_4^s]=z-2\ , \\
&[B_0^v]=[B_2^v]=d, \ [B_1^v]=z+d-2\ ,\\
&[C_0^s]=[C_2^s]=d-4(z-1), \ [C_1^s]=d+2-3z,\ [C_3^s]=d-2(z-1),\ [C_4^s]=-z\ ,\\
&[C_0^v]=[C_2^v]=d-2(z-1), \ [C_1^v]=d-z\ ,\\
&[D_0^s]=[D_2^s]=d-4(z-1),\ [D_1^s]=d-3z+2,\ [D_3^s]=d-2(z-1),\ [D_4^s]=-z\ ,\\
&[E_0^s]=[E_2^s]=d-2(z-1),\ [E_1^s]=d-z,\ [E_3^s]=d,\ [E_4^s]=z-2\ ,\\
&[D_0^v]=[D_2^v]=d-2(z-1),\ [D_1^v]=d-z,\ [\eta]=d\ .
\end{split}
\end{align}
The dependence on the thermodynamic variables of the coefficients is then partially fixed to be of the form \eqref{scaledependenceNR}. 

Using \eqref{eq:acc} in the non-relativistic limit we find that to leading order the acceleration is given by:
\begin{equation}
a^i\simeq \frac{1}{c^2} D_t v^i \simeq -\frac{1}{c^2}\left[\partial^i\left(\mu+\frac{w^2}{2} \right) +\frac{w^i}{\rho_n}\left[D_t\rho_s +\rho_s\partial_k v^k+\partial_k(\rho_s w^k)\right]+\frac{Ts}{\rho_n}\frac{\partial_i T}{T}\right]\ .
\end{equation}
Note, that the qualitatively new terms involve the gradient of the chemical potential together with the time derivative of the superfluid density. Using the equations of motion at the ideal order, an alternative choice of independent scalars and vectors could be to take:
\begin{align}
\begin{split}
\tilde S_0 &=w^i\partial_i\left(\mu+\frac{w^2}{2} \right)+\frac{w^2}{\rho_n} D_t\rho_s\ ,\\
\tilde W_0^i &= P^{ij}_w \partial_j\left(\mu+\frac{w^2}{2} \right) \ . \label{twoterms}
\end{split}
\end{align}

\section{Discussion}\label{discus}

In this work we constructed the first order hydrodynamics of quantum critical points with Lifshitz scaling and a spontaneously broken symmetry.
In the time-reversal invariant case, we found eight new parity even transport coefficients associated with flows proportional to the normal-fluid acceleration. In fact we encountered two new basic terms ((\ref{twoterms}) in the non-relativistic case). One is associated with the projection of the normal-fluid acceleration in the direction of the superfluid velocity and the second one with its transverse part. These structures were not present in the case of a superfluid without broken boost invariance.

One can think of setups in which acceleration related transport terms can be measured.
An acceleration can be induced by chemical potential difference, e.g. the flow generated between two different height levels of a fluid  in an open dam.
For an accelerating normal fluid component in a background superfluid velocity, an anisotropy would be generated in the heat flow between the parallel and the perpendicular directions with respect to the superfluid velocity.
We find it, however, difficult in general to disentangle this anisotropic contribution from the shear one induced by $S_2$ and $W_2^i$, which would vanish only if the acceleration is homogeneous throughout the fluid.
Nevertheless, we can use other setups to measure the normal fluid viscosity (such as the drag imposed on a rotating cylinder in a superfluid vessel). Subtracting this effect, we should be able to see the anisotropic heat flow due to acceleration.

The generalization of the analysis to a superconductor is done by adding the external gauge field to the
condensate phase gradient.
We suspect that the new terms depending on the acceleration may be attenuated in real (solid) superconductors due to drag terms.
One can apply an alternating electric field to the superconductor in order to excite such terms. However, since the superconducting flow exhibits zero electric resistivity, it is possible that, for frequencies where the hydrodynamic approximation is valid, the normal flow (of finite resistance) would be negligible and therefore a non-zero acceleration cannot be generated.

An important generalization of our work that we intend to study in the future is the study of time reversal and
parity violating effects.  We expect such new effects  (in addition to e.g. the time reversal breaking effects of section 3.1 of \cite{Bhattacharyya:2012xi} and  the chiral electric effect of \cite{Neiman:2011mj}),
in the hydrodynamics of parity/time reversal breaking Lifshitz superfluids and superconductors.

\section{Acknowledgements}
We would like to thank Bom Soo Kim for useful discussions.
This work is supported in part by the Israeli Science Foundation Center of Excellence, the I-CORE program of Planning and Budgeting Committee, the Israel Science Foundation (grant number 1937/12) and by the US-Israel Binational Science Foundation. The work of S.C is partially supported by the Israel Ministry of Science and Technology.

\appendix

\section{Landau's Superfluid Hydrodynamics from Relativistic Superfluid}\label{nrlimitappendix}

The relativistic energy-momentum tensor is:
\begin{equation}
T^{\mu\nu}=(\varepsilon_n+p)u^\mu u^\nu+p\eta^{\mu\nu}+f\xi_0^\mu\xi_0^\nu+\pi^{\mu\nu}\ ,
\end{equation}
We will take the $c\to\infty$ limit. The expansion of the  components of the normal-fluid velocity  are, to leading order:
\begin{equation}
u^\mu=\left(1+\frac{v^2}{2c^2}\right) \times \left(1\ ,\frac{v^i}{c} \right)\ .
\end{equation}
The expansion of the components of the superfluid velocity is:
\begin{align}
\label{phase1} &\xi_0^0 = -c\sqrt{\frac{\rho_s}{f}}\left[1+\frac{U_s}{2\rho_s c^2}\right]\ ,\\
\label{phase2} &\xi_0^i=-\sqrt{\frac{\rho_s}{f}}\left[v_s^i+\frac{U_s^i}{\rho_s c^2}\right]\ ,
\end{align}
where $f$ itself has also an expansion in $1/c$,  $f=\rho_s(1+\delta f/c^2)$.
The condition \eqref{eq:rotat} leads to:
\begin{equation}
\partial_t v_s^i+\partial^i\left(H+\frac{U_s}{2\rho_s}-\frac{\delta f}{2}\right)=0\ ,
\end{equation}
where we have defined $\mu_{\rm diss}=H/c$. As we will see below, $U_s=\rho_s v_s^2/2$. Then, in order to recover Landau's equation for the time derivative of the superfluid velocity:
\begin{equation}\label{joseph}
\partial_t v_s^i+\partial_i \left(\frac{v_s^2}{2}+ \mu + H \right)=0\ ,
\end{equation}
we should impose:
\begin{equation}
\delta f= -\frac{v_s^2}{2}-2\mu\ .
\end{equation}

We can now evaluate the Josephson condition to leading order:
\begin{equation}
u^\mu \xi_{0\,\mu} \simeq c\left[1+\frac{1}{c^2}\left(\mu+\frac{w^2}{2} \right)\right]\ .
\end{equation}
We see that the relativistic and non-relativistic chemical potentials are related as:
\begin{equation}
\mu_{\rm rel}=c+\frac{1}{c}\left(\mu+\frac{w^2}{2} \right)\ .
\end{equation}

\subsection{Thermodynamic Relations}
In the relativistic theory the energy, pressure, etc., are functions of $T$, $\mu_{\rm rel}$ and $\xi_0=\sqrt{-\xi_0^\mu\xi_{0\,\mu}}$. They satisfy the thermodynamic relations:
\begin{align}
\begin{split}
& \varepsilon_n+p =Ts+\mu_{\rm rel} q_n\ ,\\
& d\varepsilon_n =Tds+\mu_{\rm rel} dq_n-\frac{f}{2}d\xi_0^2\ .
\end{split}
\end{align}
In order to take the non-relativistic limit consistently we do the following expansion of the normal energy and charge densities:
\begin{align}
&\varepsilon_n\simeq \rho_n c^2+U_n-\rho_n\frac{v^2}{2}\ ,\\
&q_n\simeq \rho_n c-\frac{1}{c}\left(\rho_n\frac{v^2}{2}-\mu\rho_s \right)\ ,
\end{align}
and use the expansions of the chemical potential and superfluid velocity:
\begin{align}
&\mu_{\rm rel}\simeq c+\frac{1}{c}\left(\mu+\frac{w^2}{2} \right)\ ,\\
&\xi_0^2\simeq c^2+2\mu\ ,\\
&f\simeq \rho_s\ .
\end{align}
Using these expressions it is easy to check that to order $O(c^2)$ the thermodynamic identities are trivial and to $O(1)$ they become:
\begin{align}\label{eq:thermorel}
\begin{split}
&U_n+p=Ts+\mu \rho+\rho_n\frac{w^2}{2}\ ,\\
&dU_n=Tds+\mu d\rho+\frac{w^2}{2}d\rho_n\ .
\end{split}
\end{align}
The `internal energy density' $U_n$ derived from the non-relativistic limit is not exactly the internal energy $U_0$ used in textbooks \cite{Landau,Putterman}. However, both are simply related by the shift:
\begin{equation}
U_0=U_n+\rho_n\frac{w^2}{2}\ .
\end{equation}

\subsection{Hydrodynamic Equations}
The equation for the conservation of momentum:
\begin{equation}
\frac{1}{c}\partial_t T^{0i}+\partial_k T^{ki}=0\ ,
\end{equation}
becomes, at $O(c^0)$, the Navier-Stokes equation:
\begin{equation}\label{nes}
\partial_t(\rho_n v^i+\rho_s v_s^i)+\partial_k(\rho_n v^i v^k+\rho_s v_s^i v_s^k)+\partial^ip+D^i=0\ ,
\end{equation}
where we have defined:
\begin{equation}
D^i=\partial_t \pi_{(-1)}^{0i}+\partial_k \pi_{(0)}^{ki}\ .
\end{equation}

The conservation of the energy:
\begin{equation}
\frac{1}{c}\partial_t T^{00}+\partial_i T^{i0}=0\ ,
\end{equation}
gives two equations. At $O(c)$, it is the mass current conservation:
\begin{equation}\label{mc}
\partial_t(\rho_n+\rho_s)+\partial_i(\rho_n v^i+\rho_s v_s^i)=0\ .
\end{equation}
At $O(1/c)$ it is the energy conservation equation:
\begin{equation}\label{ec}
\partial_t\left(U_n+\rho_n\frac{v^2}{2}+U_s\right)+\partial_i\left(\left(U_n+\rho_n\frac{v^2}{2}+p\right)v^i+U^i_s+\frac{U_s}{2}v_s^i \right)+D_0=0\ ,
\end{equation}
where we have defined:
\begin{equation}
D_0=\partial_t \pi_{(0)}^{00}+\partial_i \pi_{(1)}^{i0}\ .
\end{equation}

We will now combine \eqref{ec} $-v \cdot$ \eqref{nes} $+\frac{v^2}{2}$\eqref{mc}. We will also set:
\begin{equation}
U_s=\rho_s \frac{v_s^2}{2}\ , \ \ U_s^i=\frac{U_s}{2}v_s^i+u_s^i\ , \ \ w^i=v_s^i-v^i\ .
\end{equation}
We get:
\begin{equation}\label{ec2}
\begin{split}
\partial_t U_n+\partial_i\left(U_n v^i+u^i_s\right)+p\partial_i v^i & +\frac{w^2}{2}(\partial_t \rho_s+\partial_i(\rho_s v_s^i))
\\
&+\rho_s w^i\left[\partial_t v_s^i+\partial_k v_s^i v_s^k \right]
+D_0-v^iD_i=0\ .
\end{split}
\end{equation}
We can now use:\footnote{This is a consequence of the curl free condition \eqref{eq:rotat}.}
\begin{equation}
\partial_k v_{s\,i}=\partial_i v_{s\,k}\ ,
\end{equation}
and combine \eqref{ec2} $-\left(\mu+\frac{w^2}{2} \right)$\eqref{mc}.\footnote{This corresponds to the relativistic combination $\partial_\mu T^{\mu\nu} u_\nu+\mu \partial_\mu J^\mu$ leading to entropy production.}
We also set:
\begin{equation}
u_s^i=(\mu+H)\rho_s w^i\ .
\end{equation}
This leads to the equation:
\begin{equation}\label{ec3}
\begin{split}
\partial_t U_n&+\partial_i\left(U_n v^i\right)+\left(p-\rho_n\frac{w^2}{2}-\mu\rho\right)\partial_i v^i-\mu(\partial_t\rho+v^i\partial_i\rho)-\frac{w^2}{2}(\partial_t \rho_n+v^i\partial_i \rho_n )
\\
&+H\partial_i(\rho_s w^i)
+\rho_s w^i\left[\partial_t v_s^i+\partial_i \frac{v_s^2}{2}+\partial_i (\mu+H)  \right]+D_0-v^iD_i=0 \ ,
\end{split}
\end{equation}
where $\rho\equiv\rho_n+\rho_s$ is the total mass density.
We now use the thermodynamic relations \eqref{eq:thermorel}
and derive the conservation of the entropy current to ideal order by using the equation for the time derivative of the superfluid velocity \eqref{joseph}.

Including dissipative terms, the entropy equation is:
\begin{equation}
T(\partial_t s+\partial_i(s v^i))+D_0-v^iD_i +H\partial_i(\rho_s w^i)=0\ .
\end{equation}
More explicitly: 
\begin{equation}
T(\partial_t s+\partial_i(s v^i))+\partial_t\pi_{(0)}^{00}+\partial_i\pi_{(1)}^{i0}-v_i(\partial_t \pi^{0i}_{(-1)}+\partial_k\pi_{(0)}^{ki}) +H\partial_i(\rho_s w^i)  =0\ .
\end{equation}
We have used that $\pi_{(0)}^{i0}=0$ and allowed for a term $O(c)$ in $\pi^{0i}$ (this is clearly only possible if we have asymmetric terms). Using now:
\begin{equation}
\partial_\mu X=T\partial_\mu\left(\frac{X}{T}\right)+X\frac{\partial_\mu T}{T}\ ,
\end{equation}
we can rewrite the entropy equation as:
\begin{align}
\begin{split}
&T\left[\partial_t \left(s+\frac{1}{T}\left(\pi_{(0)}^{00}-\pi^{0i}_{(-1)}v_i\right)\right)+\partial_i\left(s v^i+\frac{1}{T}\left(\pi_{(1)}^{i0}-\pi_{(0)}^{ik} v_k\right)\right)\right]\\
&+\left(\pi_{(0)}^{00}-\pi^{0i}_{(-1)}v_i\right)\frac{\partial_t T}{T}+\left(\pi_{(1)}^{i0}-\pi_{(0)}^{ik} v_k\right)\frac{\partial_i T}{T} +\pi_{(-1)}^{0i}\partial_t v_i+\pi_{(0)}^{ki}\partial_k v_i\\
&  +H\partial_i(\rho_s w^i) =0\ .
\end{split}
\end{align}
We will now use the frame condition in the non-relativistic limit:
\begin{equation}
0=\pi^{\mu\nu} u_\mu u_\nu \simeq \pi^{00}_{(0)}-\pi^{0i}_{(-1)}v_i +O(1/c)\ .
\end{equation}
The entropy equation then becomes:
\begin{align}\label{entropyNRap}
\begin{split}
&T\left[\partial_t s+\partial_i\left(s v^i+\frac{1}{T}\left(\pi_{(1)}^{i0}-\pi_{(0)}^{ik} v_k\right) \right)\right]\\
&+\left(\pi_{(1)}^{i0}-\pi_{(0)}^{ik} v_k  \right)\frac{\partial_i T}{T} +\pi_{(-1)}^{0i}\partial_t v_i+\pi_{(0)}^{ki}\partial_k v_i
\\ & +H \del_i(\rho_s w^i)=0\ .
\end{split}
\end{align}


\subsection{Terms Depending on the Acceleration}

From the derivation above we see that in the Lifshitz superfluid new dissipative terms proportional to the acceleration $D_t v^i$ are allowed.

One can use the ideal equations of motion of the relativistic superfluid to derive the relation \eqref{eq:acc} between the acceleration and the gradients of temperature and chemical potential that is valid to leading order in the derivative expansion.
In the non-relativistic limit, the expansion of the gradient terms is to leading order:
\begin{equation}
\begin{split}
P^{\mu\sigma} & \left[
\frac{\partial_\sigma T}{T}+
\frac{q_n T}{\varepsilon_n + p}\del_\sigma \frac{\mu}{T}  + \zeta_\sigma \frac{\del_\alpha\left(f\xi_0^\alpha\right)}{\varepsilon_n + p}
\right]
\\
&\simeq \frac{\delta^{\mu i}}{c^2}\left[\partial_i\left(\mu+\frac{w^2}{2} \right) +\frac{Ts}{\rho_n}\frac{\partial_i T}{T}+\frac{w_i}{\rho_n}(\partial_t\rho_s+\partial_k(\rho_s v_s^k))\right]\ .
\end{split}
\end{equation}
Then, to leading order, the acceleration only has non-zero spatial components:
\begin{equation}
a_i\simeq \frac{1}{c^2}D_tv_i=-\frac{1}{c^2}\left[\partial_i\left(\mu+\frac{w^2}{2} \right) +\frac{Ts}{\rho_n}\frac{\partial_i T}{T}+w_i\frac{D_t\rho_s+\rho_s\partial_k v^k+\partial_k(\rho_s w^k)}{\rho_n}\right]\ .
\end{equation}
Therefore, in this approximation we can trade the terms proportional to the acceleration by terms depending on gradients of the chemical potential.

\end{document}